\documentclass[10pt,a4paper,onecolumn]{article}
\usepackage{marginnote}
\usepackage{graphicx}
\usepackage{xcolor}
\usepackage{authblk,etoolbox}
\usepackage{titlesec}
\usepackage{calc}
\usepackage{tikz}
\usepackage{hyperref}
\hypersetup{colorlinks,breaklinks=true,
            urlcolor=[rgb]{0.0, 0.5, 1.0},
            linkcolor=[rgb]{0.0, 0.5, 1.0}}
\usepackage{caption}
\usepackage{tcolorbox}
\usepackage{amssymb,amsmath}
\usepackage{ifxetex,ifluatex}
\usepackage{seqsplit}
\usepackage{xstring}

\usepackage{float}
\let\origfigure\figure
\let\endorigfigure\endfigure
\renewenvironment{figure}[1][2] {
    \expandafter\origfigure\expandafter[H]
} {
    \endorigfigure
}

\usepackage{fixltx2e} % provides \textsubscript
\usepackage[
  backend=biber,
]{biblatex}
\bibliography{paper.bib}

\let\textttOrig=\texttt
\def\texttt#1{\expandafter\textttOrig{\seqsplit{#1}}}
\renewcommand{\seqinsert}{\ifmmode
  \allowbreak
  \else\penalty6000\hspace{0pt plus 0.02em}\fi}

\makeatletter
\let\href@Orig=\href
\def\href@Urllike#1#2{\href@Orig{#1}{\begingroup
    \def\Url@String{#2}\Url@FormatString
    \endgroup}}
\def\href@Notdoi#1#2{\def\tempa{#1}\def\tempb{#2}%
  \ifx\tempa\tempb\relax\href@Urllike{#1}{#2}\else
  \href@Orig{#1}{#2}\fi}
\def\href#1#2{%
  \IfBeginWith{#1}{https://doi.org}%
  {\href@Urllike{#1}{#2}}{\href@Notdoi{#1}{#2}}}
\makeatother

\newlength{\cslhangindent}
\newlength{\csllabelwidth}
\newenvironment{CSLReferences}[3] % #1 hanging-ident, #2 entry spacing
 {% don't indent paragraphs
  \setlength{\parindent}{0pt}
  \ifodd #1 \everypar{\setlength{\hangindent}{\cslhangindent}}\ignorespaces\fi
  \ifnum #2 > 0
  \setlength{\parskip}{#2\baselineskip}
  \fi
 }%
 {}
\usepackage{calc}

\usepackage[top=3.5cm, bottom=3cm, right=1.5cm, left=1.0cm,
            headheight=2.2cm, reversemp, includemp, marginparwidth=4.5cm]{geometry}

\titleformat{\section}
  {\normalfont\sffamily\Large\bfseries}
  {}{0pt}{}
\titleformat{\subsection}
  {\normalfont\sffamily\large\bfseries}
  {}{0pt}{}
\titleformat{\subsubsection}
  {\normalfont\sffamily\bfseries}
  {}{0pt}{}
\titleformat*{\paragraph}
  {\sffamily\normalsize}

\usepackage{fancyhdr}
\makeatletter
\let\ps@plain\ps@fancy
\fancyheadoffset[L]{4.5cm}
\fancyfootoffset[L]{4.5cm}

\definecolor{linky}{rgb}{0.0, 0.5, 1.0}

\newtcolorbox{repobox}
   {colback=red, colframe=red!75!black,
     boxrule=0.5pt, arc=2pt, left=6pt, right=6pt, top=3pt, bottom=3pt}

\newcommand{\ExternalLink}{%
   \tikz[x=1.2ex, y=1.2ex, baseline=-0.05ex]{%
       \begin{scope}[x=1ex, y=1ex]
           \clip (-0.1,-0.1)
               --++ (-0, 1.2)
               --++ (0.6, 0)
               --++ (0, -0.6)
               --++ (0.6, 0)
               --++ (0, -1);
           \path[draw,
               line width = 0.5,
               rounded corners=0.5]
               (0,0) rectangle (1,1);
       \end{scope}
       \path[draw, line width = 0.5] (0.5, 0.5)
           -- (1, 1);
       \path[draw, line width = 0.5] (0.6, 1)
           -- (1, 1) -- (1, 0.6);
       }
   }

\patchcmd{\@maketitle}{center}{flushleft}{}{}
\patchcmd{\@maketitle}{center}{flushleft}{}{}
\patchcmd{\@maketitle}{\LARGE}{\LARGE\sffamily}{}{}
\def\maketitle{{%
  
  \AB@maketitle}}
\makeatletter
\renewcommand\AB@affilsepx{ \protect\Affilfont}
\renewcommand\AB@affilnote[1]{{\bfseries #1}\hspace{3pt}}
\renewcommand{\affil}[2][]%
   {\newaffiltrue\let\AB@blk@and\AB@pand
      \if\relax#1\relax\def\AB@note{\AB@thenote}\else\def\AB@note{#1}%
        \setcounter{Maxaffil}{0}\fi
        \begingroup
        \let\href=\href@Orig
        \let\texttt=\textttOrig
        \let\protect\@unexpandable@protect
        \def\thanks{\protect\thanks}\def\footnote{\protect\footnote}%
        \@temptokena=\expandafter{\AB@authors}%
        {\def\\{\protect\\\protect\Affilfont}\xdef\AB@temp{#2}}%
         \xdef\AB@authors{\the\@temptokena\AB@las\AB@au@str
         \protect\\[\affilsep]\protect\Affilfont\AB@temp}%
         \gdef\AB@las{}\gdef\AB@au@str{}%
        {\def\\{, \ignorespaces}\xdef\AB@temp{#2}}%
        \@temptokena=\expandafter{\AB@affillist}%
        \xdef\AB@affillist{\the\@temptokena \AB@affilsep
          \AB@affilnote{\AB@note}\protect\Affilfont\AB@temp}%
      \endgroup
       \let\AB@affilsep\AB@affilsepx
}
\makeatother

\renewcommand\Affilfont{\sffamily\small\mdseries}
\ifnum 0\ifxetex 1\fi\ifluatex 1\fi=0 % if pdftex
  \usepackage[T1]{fontenc}
  \usepackage[utf8]{inputenc}

\else % if luatex or xelatex
  \ifxetex
    \usepackage{mathspec}
    \usepackage{fontspec}

  \else
    \usepackage{fontspec}
  \fi
  \defaultfontfeatures{Ligatures=TeX,Scale=MatchLowercase}

\fi
\IfFileExists{upquote.sty}{\usepackage{upquote}}{}
\IfFileExists{microtype.sty}{%
\usepackage{microtype}
\UseMicrotypeSet[protrusion]{basicmath} % disable protrusion for tt fonts
}{}

\usepackage{hyperref}
\hypersetup{unicode=true,
            pdftitle={QuasinormalModes.jl: A Julia package for computing discrete eigenvalues of second order ODEs},
            pdfborder={0 0 0},
            breaklinks=true}
\let\addcontentslineOrig=\addcontentsline
\def\addcontentsline#1#2#3{\bgroup
  \let\texttt=\textttOrig\addcontentslineOrig{#1}{#2}{#3}\egroup}
\let\markbothOrig\markboth
\def\markboth#1#2{\bgroup
  \let\texttt=\textttOrig\markbothOrig{#1}{#2}\egroup}
\let\markrightOrig\markright
\def\markright#1{\bgroup
  \let\texttt=\textttOrig\markrightOrig{#1}\egroup}

\usepackage{graphicx,grffile}
\makeatletter
\def\maxwidth{\ifdim\Gin@nat@width>\linewidth\linewidth\else\Gin@nat@width\fi}
\def\maxheight{\ifdim\Gin@nat@height>\textheight\textheight\else\Gin@nat@height\fi}
\makeatother
\setkeys{Gin}{width=\maxwidth,height=\maxheight,keepaspectratio}
\IfFileExists{parskip.sty}{%
\usepackage{parskip}
}{% else
\setlength{\parindent}{0pt}
\setlength{\parskip}{6pt plus 2pt minus 1pt}
}
\ifx\paragraph\undefined\else
\let\oldparagraph\paragraph
\renewcommand{\paragraph}[1]{\oldparagraph{#1}\mbox{}}
\fi
\ifx\subparagraph\undefined\else
\let\oldsubparagraph\subparagraph
\renewcommand{\subparagraph}[1]{\oldsubparagraph{#1}\mbox{}}
\fi

\title{\texttt{QuasinormalModes.jl}: A Julia package for computing
discrete eigenvalues of second order ODEs}

        \author[1]{Lucas Timotheo Sanches}
    
      \affil[1]{Centro de Ciências Naturais e Humanas, Universidade
Federal do ABC (UFABC)}
  \date{\vspace{-7ex}}

\begin{document}
\maketitle

\marginpar{

  \begin{flushleft}
  %\hrule
  \sffamily\small

  {\bfseries DOI:} \href{https://doi.org/10.21105/joss.04077}{\color{linky}{10.21105/joss.04077}}

  \vspace{2mm}

  {\bfseries Software}
  \begin{itemize}
    \setlength\itemsep{0em}
    \item \href{https://github.com/openjournals/joss-reviews/issues/4077}{\color{linky}{Review}} \ExternalLink
    \item \href{https://github.com/lucass-carneiro/QuasinormalModes.jl/}{\color{linky}{Repository}} \ExternalLink
    \item \href{https://doi.org/10.5281/zenodo.6478503}{\color{linky}{Archive}} \ExternalLink
  \end{itemize}

  \vspace{2mm}

  \par\noindent\hrulefill\par

  \vspace{2mm}

  {\bfseries Editor:} \href{http://pdebuyl.be/}{Pierre de Buyl} \ExternalLink \\
  \vspace{1mm}
    {\bfseries Reviewers:}
  \begin{itemize}
  \setlength\itemsep{0em}
    \item \href{https://github.com/JamieBamber}{@JamieBamber}
    \item \href{https://github.com/cescalara}{@cescalara}
    \end{itemize}
    \vspace{2mm}

  {\bfseries Submitted:} 22 September 2021\\
  {\bfseries Published:} 25 May 2022

  \vspace{2mm}
  {\bfseries License}\\
  Authors of papers retain copyright and release the work under a Creative Commons Attribution 4.0 International License (\href{http://creativecommons.org/licenses/by/4.0/}{\color{linky}{CC BY 4.0}}).

  \end{flushleft}
}

\hypertarget{summary}{%
\section{Summary}\label{summary}}

In General Relativity, when perturbing a black hole with an external
field, or particle, the system relaxes by emitting damped gravitational
waves, known as \emph{quasinormal modes}. These are the characteristic
frequencies of the black hole and gain the \emph{quasi} prefix due to
the fact that they have a real frequency, which represents the
oscillation frequency of the response, and an imaginary frequency that
represents the decay rate of said oscillations. In many cases, such
perturbations can be described by a second order homogeneous ordinary
differential equation (ODE) with discrete complex eigenvalues.

Determining these characteristic frequencies quickly and accurately for
a large range of models is important for many practical reasons. It has
been shown that the gravitational wave signal emitted at the final stage
of the coalescence of two compact objects is well described by
quasinormal modes (Buonanno et al., 2007; Seidel, 2004). This means that
if one has access to a database of quasinormal modes and of
gravitational wave signals from astrophysical collision events, it is
possible to characterize the remnant object using its quasinormal
frequencies. Since there are many different models that aim to describe
remnants, being able to compute the quasinormal frequencies for such
models in a reliable way is paramount for confirming or discarding them.

\hypertarget{statement-of-need}{%
\section{Statement of need}\label{statement-of-need}}

\texttt{QuasinormalModes.jl} is a \texttt{Julia} package for computing
the quasinormal modes of any General Relativity model whose perturbation
equation can be expressed as second order homogeneous ODE. Not only
that, the package can be used to compute the discrete eigenvalues of
\emph{any} second order homogeneous ODE (such as the energy eigenstates
of the time independent Schrödinger equation) provided that these
eigenvalues actually exist. The package features a flexible and user
friendly API where the user simply needs to provide the coefficients of
the problem ODE after incorporating boundary and asymptotic conditions
on it. The user can also choose to use machine or arbitrary precision
arithmetic for the underlying floating point operations involved and
whether or not to do computations sequentially or in parallel using
threads. The API also tries not to force any particular workflow on the
users so that they can incorporate and adapt the existing functionality
on their research pipelines without unwanted intrusions. Often user
friendliness, flexibility and performance are treated as mutually
exclusive, particularly in scientific applications. By using
\texttt{Julia} as an implementation language, the package can have all
of theses features simultaneously.

Another important motivation for using \texttt{Julia} and writing this
package was the lack of generalist, free (both in the financial and
license-wise sense) open source tools that serve the same purpose. More
precisely, there are tools which are free and open source, but run on
top of a proprietary paid and expensive software framework such as the
ones developped by Jansen (2017) and Fortuna \& Vega (2020), which are
both excellent packages that aim to perform the same task as
\texttt{QuasinormalModes.jl} and can be obtained and modified freely
but, unfortunately, require the user to own a license of the proprietary
\texttt{Wolfram\ Mathematica} CAS. Furthermore, their implementations
are limited to solve problems where the eigenvalues must appear in the
ODE as a polynomials of order \(p\). While this is not prohibitively
restrictive to most astrophysics problems, it can be an important
limitation in other areas. There are also packages that are free and run
on top of \texttt{Mathematica} but are not aimed at being general
eigenvalue solvers at all, such as the one by O'Toole et al. (2019),
that can only compute modes of Schwarzschild and Kerr black holes.
Finally, the Python package by Stein (2019) is open source and free but
can only compute Kerr quasinormal modes.

\texttt{QuasinormalModes.jl} fills the existing gap for free, open
source tools that are able to compute discrete eigenvalues (and in
particular, quasinormal modes) efficiently for a broad class of models
and problems. The package was developed during the author's PhD research
where it is actively used for producing novel results that shall appear
in the author's thesis. It is also actively used in a collaborative
research effort (of which the author is one of the members) for
computing quasinormal modes produced by perturbations with integer (but
different than 0) spins and semi-integer spins. These results are being
contrasted with those obtained by other methods and so far show
excellent agreement with each other and with literature results.

\hypertarget{underlying-algorithm}{%
\section{Underlying algorithm}\label{underlying-algorithm}}

\texttt{QuasinormalModes.jl} internally uses a relativity new numerical
method called the Asymptotic Iteration Method (AIM). The method was
introduced by Ciftci et al. (2003) but the actual implementation used in
this package is based on the revision performed by Cho et al. (2012).
The main purpose of the (AIM) is to solve the following general linear
homogeneous second order ODE: \begin{equation}
    y^{\prime\prime}(x) - \lambda_0(x)y^\prime(x) - s_0(x)y(x) = 0,
    \label{eq:aim_general_ode}
\end{equation} where primes denote derivatives with respect to to the
variable \(x\) (that is defined over some interval that is not
necessarily bounded), \(\lambda_0(x) \neq 0\) and
\(s_0(x) \in C_\infty\). The method is based upon the following theorem:
let \(\lambda_0\) and \(s_0\) be functions of the variable
\(x \in (a,b)\) that are \(C_\infty\) on the same interval, the solution
of the differential equation, Eq. \eqref{eq:aim_general_ode}, has the
form \begin{equation}
    y(x) = \exp\left( -\int\alpha\mathrm{d} t \right) \times \left[ C_2 + C_1 \int^{x} \exp \left( \int^{t} ( \lambda_0(\tau) + 2\alpha(\tau) )\mathrm{d} \tau \right) \mathrm{d} t \right]
    \label{eq:aim_general_solution}
\end{equation} if for some \(n>0\) the condition \begin{equation}
    \delta \equiv s_n\lambda_{n-1} - \lambda_{n}s_{n-1} = 0
    \label{eq:aim_delta_definition}
\end{equation} is satisfied, where \begin{align}
    \lambda_k(x) \equiv & \lambda^\prime_{k-1}(x) + s_{k-1}(x) + \lambda_0(x)\lambda_{k-1}(x) \label{eq:aim_lambda_k}\\
    s_k(x) \equiv & s^\prime_{k-1}(x) + s_0\lambda_{k-1}(x) \label{eq:aim_sk}
\end{align} where \(k\) is an integer that ranges from \(1\) to \(n\).

Provided that the theorem is satisfied we can find both the eigenvalues
and eigenvectors of the second order ODE using, respectively, Eq.
\eqref{eq:aim_delta_definition} and Eq. \eqref{eq:aim_general_solution}.
Due to the recursive nature of Eq.\eqref{eq:aim_lambda_k} and
Eq.\eqref{eq:aim_sk}, to compute the quantization condition,
Eq.\eqref{eq:aim_delta_definition}, using \(n\) iterations the \(n\)-th
derivatives of \(\lambda_0\) and \(s_0\) must be computed multiple
times. To address this issue, Cho et al. (2012) proposed the use of a
Taylor expansion of both \(\lambda\) and \(s\) around a point \(\xi\)
where the AIM is to be performed. This improved version is implemented
in \texttt{QuasinormalModes.jl}

\hypertarget{benchmark}{%
\section{Benchmark}\label{benchmark}}

To show \texttt{QuasinormalModes.jl} in action, this section provides a
simple benchmark where the fundamental quasinormal mode
(\(n=\ell=m=s=0\)) of a Schwarzschild black hole was computed using 16
threads on a Intel(R) Core(TM) i9-7900X @ 3.30GHz CPU with 256 bit
precision floating point numbers.

\begin{figure}
\centering
\includegraphics[width=0.6\textwidth,height=\textheight]{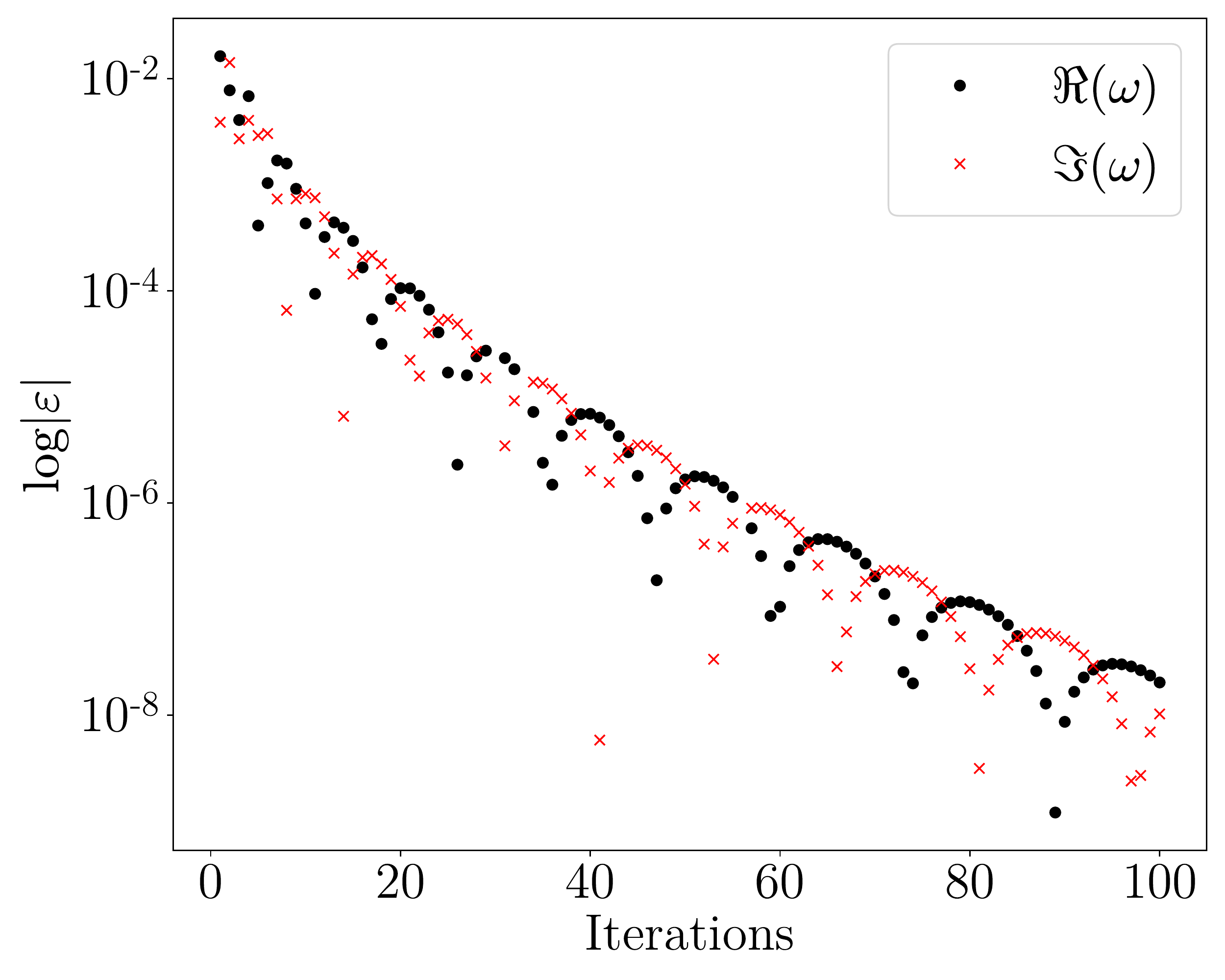}
\caption{Error convergence for the fundamental Schwarzschild quasinormal
mode as a function of the number of AIM
iterations.\label{fig:convergence}}
\end{figure}

\begin{figure}
\centering
\includegraphics[width=0.6\textwidth,height=\textheight]{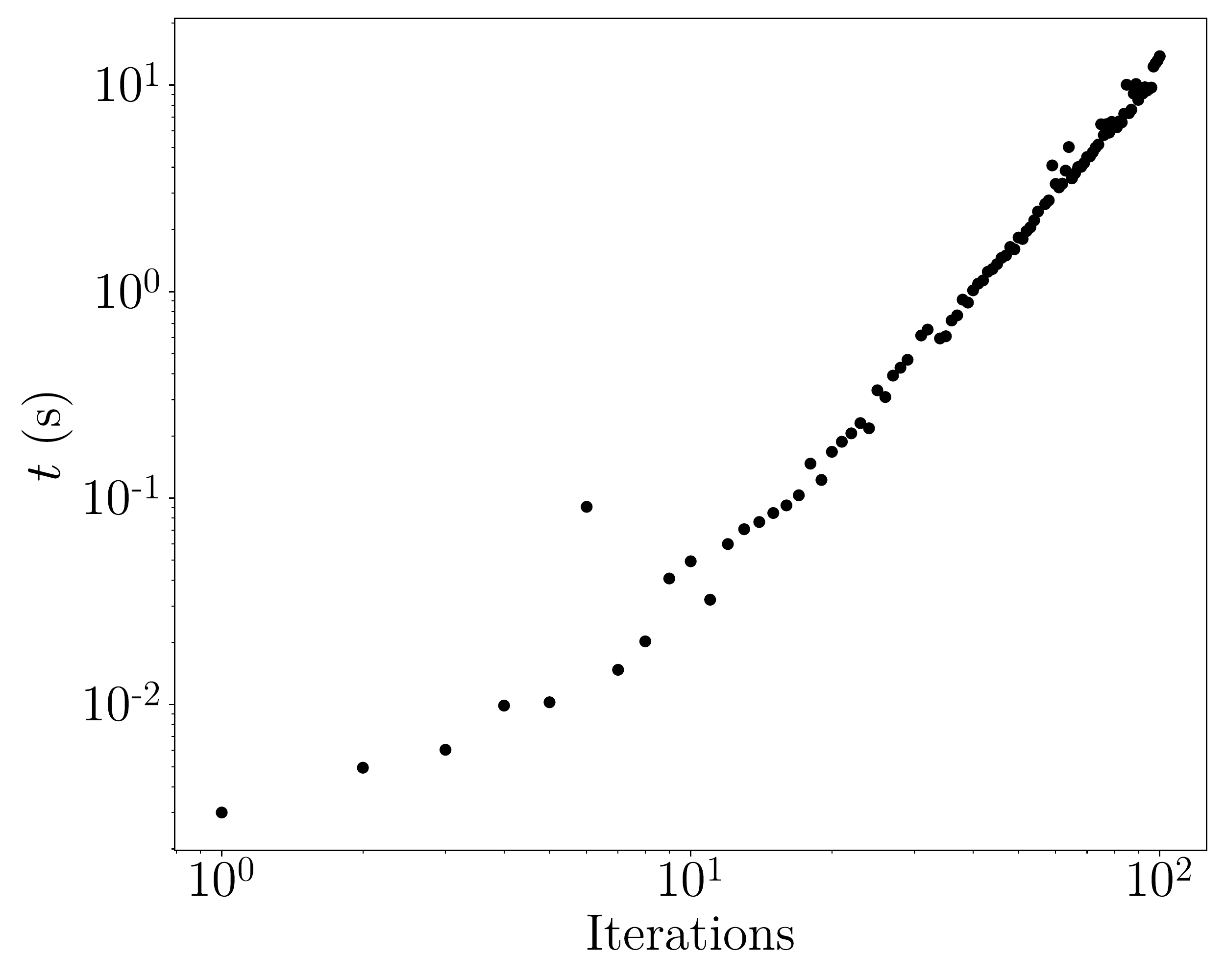}
\caption{Time taken to compute the fundamental Schwarzschild quasinormal
mode as a function of the number of AIM iterations on a logarithmic
scale.\label{fig:time}}
\end{figure}

In order to quantify the rate at which the method converges to the
correct results, the error measure \(\varepsilon\) was defined as
follows: given the computed real and imaginary quasinormal frequencies,
denoted respectively by \(\omega_R\) and \(\omega_I\), and the reference
frequencies given by Berti et al. (2009), denoted respectively as
\(\overline{\omega}_R\) and \(\overline{\omega}_I\) we have that
\(|\varepsilon| = |\omega_{R,I} - \overline{\omega}_{R,I}|\). In
\autoref{fig:convergence} \(\varepsilon\) is plotted as a function of
the number of iterations for both the real and imaginary frequencies. We
can see that, as the number of iterations increases, the error in the
computed values rapidly decreases.

In \autoref{fig:time}, the time required to perform a certain number of
iterations is plotted in logarithmic scale. Each data point is the
arithmetic mean time obtained after 10 runs of the algorithm for a given
number of iterations. The time measurement takes care to exclude the
overhead induced at ``startup'' due to Julia's JIT compilation. Even
though the time taken to perform a certain number of iterations
increases with a power law, the time scale required to achieve highly
accurate results is still around 10s. This time would be even smaller if
one choose to use built-in floating point types instead of arbitrary
precision numbers.

\hypertarget{acknowledgements}{%
\section{Acknowledgements}\label{acknowledgements}}

I would like to thank my PhD advisor, Dr.~Maurício Richartz, for
introducing me to the AIM and providing helpful discussions and comments
about the method's inner workings as well as helpful revision and
comments about the paper's structure. I would also like to thank
Dr.~Iara Ota for the helpful comments, discussions and revision of this
paper. I would also like to thank Dr.~Erik Schnetter, Soham Mukherjee
and Stamatis Vretinaris for the help and discussions regarding root
finding methods and to Dr.~Schnetter for directly contributing
documentation typo corrections and suggestions for improving the
package's overall presentation and documentation. This research was
supported by the Coordenação de Aperfeiçoamento de Pessoal de Nível
Superior (CAPES, Brazil) - Finance Code 001.

\hypertarget{references}{%
\section*{References}\label{references}}
\addcontentsline{toc}{section}{References}

\hypertarget{refs}{}
\begin{CSLReferences}{1}{0}
\leavevmode\hypertarget{ref-berti_ringdown}{}%
Berti, E., Cardoso, V., \& Starinets, A. O. (2009). {Quasinormal modes
of black holes and black branes}. \emph{Class. Quant. Grav.}, \emph{26},
163001. \url{https://doi.org/10.1088/0264-9381/26/16/163001}

\leavevmode\hypertarget{ref-buonanno}{}%
Buonanno, A., Cook, G. B., \& Pretorius, F. (2007). {Inspiral, merger
and ring-down of equal-mass black-hole binaries}. \emph{Phys. Rev. D},
\emph{75}, 124018. \url{https://doi.org/10.1103/PhysRevD.75.124018}

\leavevmode\hypertarget{ref-aim_improved}{}%
Cho, H. T., Cornell, A. S., Doukas, J., Huang, T. R., \& Naylor, W.
(2012). {A New Approach to Black Hole Quasinormal Modes: A Review of the
Asymptotic Iteration Method}. \emph{Adv. Math. Phys.}, \emph{2012},
281705. \url{https://doi.org/10.1155/2012/281705}

\leavevmode\hypertarget{ref-aim_original}{}%
Ciftci, H., Hall, R. L., \& Saad, N. (2003). Asymptotic iteration method
for eigenvalue problems. \emph{Journal of Physics A: Mathematical and
General}, \emph{36}(47), 11807--11816.
\url{https://doi.org/10.1088/0305-4470/36/47/008}

\leavevmode\hypertarget{ref-spectralbp}{}%
Fortuna, S., \& Vega, I. (2020). \emph{{Bernstein spectral method for
quasinormal modes and other eigenvalue problems}}.
\url{http://arxiv.org/abs/2003.06232}

\leavevmode\hypertarget{ref-qnmspectral}{}%
Jansen, A. (2017). {Overdamped modes in Schwarzschild-de Sitter and a
Mathematica package for the numerical computation of quasinormal modes}.
\emph{Eur. Phys. J. Plus}, \emph{132}(12), 546.
\url{https://doi.org/10.1140/epjp/i2017-11825-9}

\leavevmode\hypertarget{ref-bhpt_quasinormalmodes}{}%
O'Toole, C., Macedo, R., Stratton, T., \& Wardell, B. (2019).
QuasiNormalModes. In \emph{GitHub repository}. GitHub.
\url{https://github.com/BlackHolePerturbationToolkit/QuasiNormalModes}

\leavevmode\hypertarget{ref-seidel}{}%
Seidel, E. (2004). Nonlinear impact of perturbation theory on numerical
relativity. \emph{Classical and Quantum Gravity}, \emph{21}(3),
S339--S349. \url{https://doi.org/10.1088/0264-9381/21/3/021}

\leavevmode\hypertarget{ref-bhpt_qnm}{}%
Stein, L. C. (2019). {qnm: A Python package for calculating Kerr
quasinormal modes, separation constants, and spherical-spheroidal mixing
coefficients}. \emph{Journal of Open Source Software}, \emph{4}(42),
1683. \url{https://doi.org/10.21105/joss.01683}

\end{CSLReferences}

\end{document}